# Water-based and Biocompatible 2D Crystal Inks: from Ink Formulation to All-Inkjet Printed Heterostructures

D. McManus[1], S. Vranic[2], F. Withers[3], V. Sanchez-Romaguera[4], M. Macucci[5], H. Yang[1], R. Sorrentino[1], K. Parvez[1], S. Son[1], G. Iannaccone[5], K. Kostarelos[2], G. Fiori[5], C. Casiraghi[1*]

1 School of Chemistry, University of Manchester, UK

2 Nanomedicine Lab, Faculty of Medical and Human Sciences, University of Manchester, UK

3 School of Physics and Astronomy, University of Manchester, UK

4 Manchester Enterprise Centre, Alliance Manchester Business School, University of Manchester, UK

5 Dipartimento di Ingegneria dell'Informazione, Universita' di Pisa, Pisa, Italy

* Corresponding author email:  cinzia.casiraghi@manchester.ac.uk

**Fully exploiting the properties of 2D crystals requires a mass production method able to produce heterostructures of arbitrary complexity on any substrate, including plastic. Solution processing of graphene allows simple and low-cost techniques such as inkjet printing to be used for device fabrication. However, available inkjet printable formulations are still far from ideal as they are either based on toxic solvents, have low concentration, or require time-consuming and expensive formulation processing. In addition, none of those formulations are suitable for thin-film heterostructure fabrication due to the re-mixing of different 2D crystals, giving rise to uncontrolled interfaces, which results in poor device performance and lack of reproducibility.**

**In this work we show a general formulation engineering approach to achieve highly concentrated, and inkjet printable water-based 2D crystal formulations, which also provides optimal film formation for multi-stack fabrication. We show examples of all-inkjet printed heterostructures, such as large area arrays of photosensors on plastic and paper and programmable logic memory devices, fully exploiting the design flexibility of inkjet printing. Finally, dose-escalation cytotoxicity assays in vitro also confirm the inks biocompatible character, revealing the possibility of extending use of such 2D crystal formulations to drug delivery and biomedical applications.**

The electronics industry has been dominated by metals and complementary metal-oxide-semiconductor (CMOS) technology. However, constraints related to materials choice clearly appear in transparent and flexible electronics, heat management and rapid customisation – all of which present challenges to traditional fabrication methods. An important advance was obtained with the introduction of conductive polymers,[1] which allow simple, versatile, and low-cost techniques, such as inkjet printing, to be used for manufacturing functional devices.[2-4]

The isolation of graphene[5] has unveiled a wide range of novel 2-Dimensional (2D) materials with outstanding properties.[6-8] This new class of materials shows great promise for use in flexible electronics because their atomic thickness allows for maximum electrostatic control, optical transparency, sensitivity and mechanical flexibility.[6,9] In addition, since 2D crystals are characterized by out of plane Van der Waals interactions, they can be easily combined in one multi-layer stack, offering unprecedented control of the properties and functionalities of the resulting heterostructure-based device.[10] In this framework, inkjet printing can provide a very attractive route to low-cost and large-scale fabrication of heterostructures on any substrate. Furthermore, inkjet printing allows fabrication of complex heterostructures, which can provide multiple functionalities and improved performance.[11] Methods such as vacuum filtration and spin/spray coating, which have been previously used for heterostructure fabrication,[12,13] offer poor control of thickness and roughness of the layers, and have very limited design flexibility, in particular for fabrication of complex heterostructures and arrays.

To make inkjet printing a suitable technique for fabrication of all-printed heterostructures, it is necessary to carefully engineer 2D-crystal inks. Available inkjet printable formulations, produced by Liquid-Phase Exfoliation (LPE),[14] are still far from ideal as they are either based on toxic and expensive solvents,[15-17] or require time-consuming and expensive formulation processing,[18-22] substrate functionalization[17] or need relatively high temperature to dry,[16,17] which limits the range of substrates that can be used. In addition, none of those formulations are suitable for thin-film heterostructure fabrication, which requires multi-stack formation with well defined interfaces. Fully printing a multi-layer stack is a very well known challenge for printing technology[3,23]: the different materials in the stack tend to re-disperse at the interface, producing uncontrolled interfaces, (Supporting Information, e.g. Figure S2), resulting in poor device reproducibility and performance. Better control of interfacial effects and processing conditions allowed significant improvements in the performance of organic field effect transistors.[24]

Making new printable formulations of functional materials is very challenging: inkjet printing requires the ink to have specific physical properties[3,4]; water, for example, is unsuitable for both LPE[14] and inkjet printing.[4] Surfactants can be added to water to both lower the surface tension and to stabilise the exfoliated nanosheets.[25] Such inks are not ideal for printed electronics as they have a low concentration of 2D crystals (<0.1 weight %)[26] and contain a high amount of residual surfactant. Other examples of printable water-based inks refer to graphene oxide (GO) or reduced-GO, which are defective and require thermal or chemical treatments.

In this work we developed a simple method for the production of highly concentrated, stable, inkjet printable, water-based inks that can be formulated for a range of 2D materials. No solvent exchange, chemical treatment or harsh conditions are used. The ink composition has been optimized to achieve optimal film formation for multi-stack formation, allowing fabrication of *all-inkjet printed heterostructures*, such as arrays of all-printed photosensors over $cm^2$ areas on plastic and programmable logic devices completely made of 2D-crystals. Because the inks are expected to find utility in several applications as different consumer products, we also investigated possible adverse effects from exposure and determined their safety limitations by performing dose-escalation cytotoxicity assays *in vitro* using lung and skin cells. Overall, no significant cytotoxic responses compared to untreated cells were observed for all doses. Therefore, our inks could find important applications also in drug delivery and biomedical applications.

Amongst steric and electrostatic stabilisers, pyrene sulfonic acid derivatives are able to produce concentrated, stable (up to 1 year) water-based 2D crystal dispersions,[27-29] characterized by a high graphene to stabiliser ratio, compared to surfactants.[30] This exfoliation method has been shown to produce high quality (i.e. without oxygen-based groups) graphene dispersions.[27] However, these formulations are not inkjet printable (Figure S1-a in Supplementary Information). In order to be printable, inks must have a viscosity ($\eta$), surface tension ($\gamma$) and density ($\rho$) within certain ranges for a set nozzle diameter ($\alpha$). The inverse Ohnesorge number is commonly used to predict if an ink will form stable drops: $Z = \sqrt{\gamma\rho\alpha}/\eta$.[31] The ink will produce stable drops if 1 < Z < 14.[31] For a nozzle diameter of 21.5μm, Z is ≈ 40 for water. In order to lower Z, the surface tension must be reduced and the viscosity increased using additives. Here triton x-100 was used to decrease the surface tension from ≈73 mN m$^{-1}$ to ≈40 mN m$^{-1}$, while propylene glycol was used as co-solvent to increase the viscosity from 1 mPa·s to 1.37 mPa·s. This also disrupts the weak Marangoni flow, which helps reduce the coffee-ring effect.[32] These values of surface tension and viscosity give Z ≈20 for the modified water-based ink. Despite Z >14, the drop is stable, and no satellite drops or nozzle blocking are observed (Video 1 in Supplementary Information). This is in agreement with water/ethylene glycol inks,[33] which can be printed with Z = 35.5 and N-Methyl-2-pyrrolidone (NMP)-based graphene inks, which show good printability with Z > 20.[16,17]

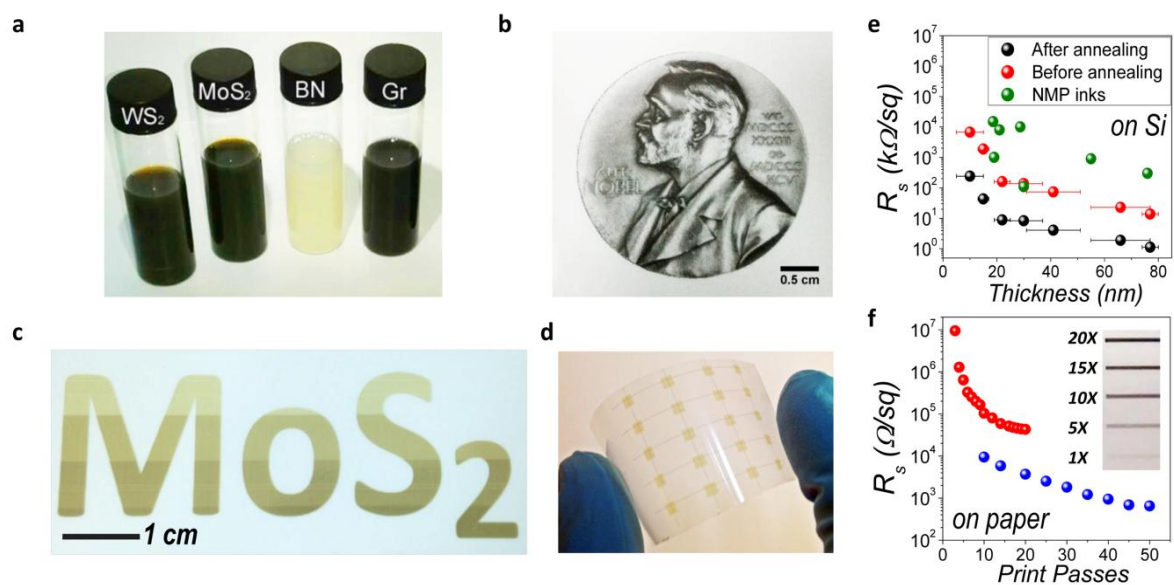

**Figure 1 | Range of inkjet printable inks and their properties. a,** Optical Image of water-based 2D-crystal inks. **b,** Nobel medal printed with water-based graphene ink on paper PEL 60 **c,** Printed "MoS$_2$" logo with water-based MoS$_2$ ink on paper PEL 60 with increasing number of printed passes (from 1 to 4, moving from top to bottom). Note the relatively good contrast with the paper obtained already with one printed pass. **d,** Optical picture of an array of all-printed Gr/WS$_2$/Gr heterostructures on PET. **e,** Sheet resistance as a function of the thickness of graphene lines (1 cm length) printed on silicon (Si/SiO$_2$), before and after annealing. Ink concentration ≈ 2 mg/mL. The data is compared with the sheet resistance reported for NMP-based graphene inks from Ref. 17. **f,** Sheet resistance for increasing number of printed passes for graphene lines (2 cm length) printed on paper PEL 60 (inset shows the optical pictures of the printed lines made with 1, 5, 10, 15 and 20 printed passes, from bottom to top) with drop spacing of 40 μm (red points). Note that in the case of paper, its porosity allows printing with small drop spacing (25 μm), leading to deposition of larger amount of material per unit area with fewer printing passes, eventually minimizing $R_s$ (blue points). Ink concentration ≈ 3 mg/mL.

Note also that the concentration of surfactant is a very important parameter, which needs to be carefully optimized (Supplementary Information, Section 1). Surfactant and co-solvent can be added before or after exfoliation of the bulk crystal. In this work we modify the dispersion after exfoliation; note that we have never observed changes in concentration and stability (i.e. re-aggregation), therefore the formulation modification does not alter the thickness distribution and quality of the flakes, as also confirmed by electrical measurements. Triton x-100, a non-ionic surfactant, was chosen on purpose to avoid disrupting the electrostatic stabilization of graphene flakes provided by pyrene sulfonic acid derivatives.

Figure 1a shows an optical picture of the inks. Several 2D materials have been successfully exfoliated and printed, including graphene, MoS$_2$, WS$_2$ and h-BN. Figure 1 b and c show some examples of printed patterns obtained with water-based graphene and MoS$_2$ inks, respectively, on PEL P60 paper. Other examples can be found in the Supplementary Information, Figure S7. We point out that no treatment of the substrate is used and the whole printing process is performed under ambient conditions. Figure 1 e and f show the values of sheet resistance $R_s$ of printed graphene lines on silicon (Si/SiO$_2$) and paper, respectively, for increasing number of printed passes: $R_s$ is infinite before the percolation threshold is reached,[34] after which $R_s$ decreases rapidly until the percolation to bulk threshold transition, at which point the graphene behaves like a bulk material, with only small changes in sheet resistance resulting from each subsequent printing pass. $R_s$ values range between ≈10 MΩ/□ and ≈1 kΩ/□, which are comparable with literature results obtained for other types of printable graphene inks.[16,17,19,35] Figure 1e compares $R_s$ values obtained with our water-based inks and NMP-based graphene inks printed on untreated silicon (Si/SiO$_2$).[17] Note that thermal annealing at 300°C for 1h under nitrogen atmosphere can be used to further lower $R_s$, Figure 1e, but this process can be used only for some substrates, such as silicon (Si/SiO$_2$), quartz, and polyimide. Note that the thickness of the printed line varies as a function of ink concentration. In figure 1(e), an ink with concentration of 2 mg/mL was used, giving rise to a thickness of ≈5 nm per printing pass on silicon (Si/SiO$_2$). In the case of paper, its porosity allows fast drying and therefore printing with reduced drop spacing. By optimizing the drop spacing for a fixed ink concentration (2-3 mg/mL), we have been able to obtain $R_s$ below 1KΩ/sq above 40 printing passes (blue points, Figure 1(f)). The thickness per pass on paper cannot be measured accurately due to the roughness of the substrate.

The basic ink formulation process can be further improved by removing excess pyrene after exfoliation with the use of a washing step. This allows us to increase graphene concentration up to 8 mg/mL (0.8 w%), enabling graphene conductive lines to be printed with just one printing pass. This is highly favourable for inkjet printing as a single pass printing allows reducing fabrication time and costs. This is visible also in figure 1c, where just one printed pass gives rise to a noticeable colour contrast of the MoS$_2$ pattern to the paper.

In printing technology it is common to modify a formulation to enhance some of its properties (such as conductivity, mechanical integrity and adhesion to the substrate).[3] In order to print a heterostructure, i.e. a multi-layer stack, a binder was also added to the printable ink to minimize the re-dispersion at the interface. The type and amount of binder depends on many factors including the concentration, solvent and stability of the ink.[36] In our case, we selected polysaccharide Xanthan gum (XG) because it is soluble in water and provides the dual benefits of acting as a binder and requiring a low concentration to increase the viscosity. This allows retention of the inks electrical properties. The addition of the binder produces inks with non-Newtonian viscosity, i.e. an ink with shear-thinning properties, which helps form uniform lines as the viscosity increases substantially after the droplets are deposited on the surface. A major advantage of XG over other binders is its biocompatibility.[37] Note that a change in the ink chemistry is known to affect the stability and printability of the ink; in our case, the binder gives rise to a small filament (tail) on ejection of the ink from the nozzle which re-joins with the main drop without forming a satellite drop, so the ink is still printable (Supplementary Information, Figure S8). Note that the binder can be added only after exfoliation as the high shear forces involved in exfoliation cause polymer degradation.[38]

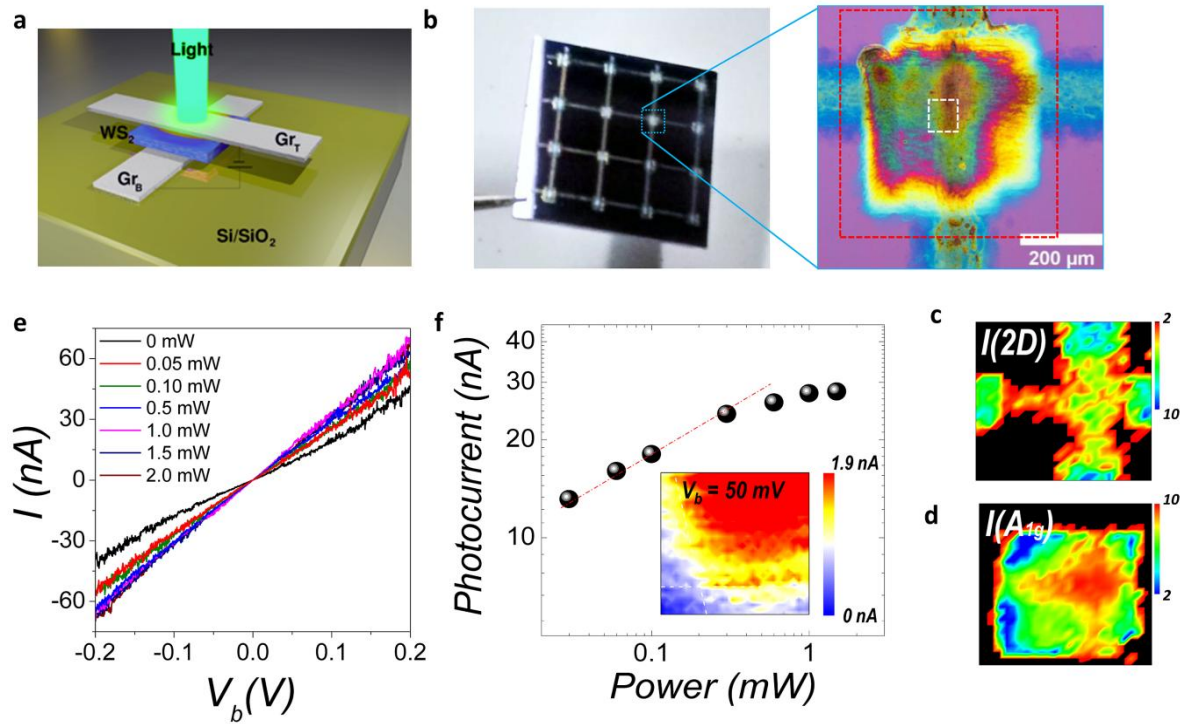

**Figure 2 | Fully inkjet printed heterostructures on Si/SiO$_2$. a,** Schematic of an all-printed Gr$_B$/WS$_2$/Gr$_T$ heterostructure on Si/SiO$_2$ substrate. **b,** (left panel) Optical picture of an array of 4 x 4 Gr$_B$/WS$_2$/Gr$_T$ heterostructures printed on Si/SiO$_2$; (right panel) Optical picture of one of the heterostructures, showing the two graphene electrodes and the square of photo-active material; **c,** Raman map measured on the red dotted square in **b** (right panel) showing the intensity of the 2D peak and the A$_{1g}$ peak (**d**). **e,** I-V$_b$ curves measured under different laser power (at 488nm); **f,** photo-current measured at V$_b$ = 200 mV as a function of the laser power; inset, photo-current map measured on the white dotted square in **b**, right panel. The red dotted line is a guide for the eyes.

With this modified water-based ink we fabricated a heterostructure-based photodetector onto Si/SiO$_2$ by printing a first graphene line of ≈50 nm thickness at 50 °C. This acts as bottom graphene electrode (Gr$_B$). A WS$_2$ square of ≈100 nm thickness, acting as photo-active element, was then printed across the graphene line and finally a second graphene line (≈50 nm thickness) was printed perpendicularly to the first, acting as top graphene electrode (Gr$_T$). The schematic of the Gr$_B$/WS$_2$/Gr$_T$ heterostructure is shown in Figure 2a. Figure 2b (left panel) shows an optical picture of an array of 16 heterostructures on an area of 1 cm x 1 cm on silicon (Si/SiO$_2$). Figure 2b (right panel) shows a high magnification optical picture of one of the heterostructures. Note the typical interference colours, which are already an indication of multiple-stack formation. Raman mapping at 488 nm excitation wavelength was performed on the dotted red square (≈500 µm x 500 µm) in Figure 2b (right panel). Figure 2 c and d show the Raman maps of the intensity of the 2D peak of graphene and the $A_{1g}$ peak of $WS_2$, respectively. This figure shows that the material has been deposited uniformly. The Raman spectra show the typical features of liquid-phase exfoliated graphene and $WS_2$ (Supplementary Information, Figure S9). After printing was completed, the sample was annealed under $N_2$ atmosphere at 300°C for 1 hour to remove residual moisture. The non-linear I-V curve, Figure 2e, confirms that multi-stacking layers were formed, with no bias voltage required for photocurrent response (Supplementary Information, Section 4). Photocurrent mapping, inset Figure 2f, shows uniform photocurrent response in the overlap region between the two graphene electrodes and the $WS_2$ square. The photo-current efficiency of the printed devices is comparable with other heterostructures fabricated from inks using non-scalable methods.[12] Figure 2f also shows a power law for photocurrent vs laser power, which breaks at laser powers above ≈0.5 mW. We found that all of the 16 heterostructures show the same I-V characteristic, within the experimental error (further data are included in the Supplementary Information, Section 4). Therefore, device fabrication yield on silicon (Si/SiO$_2$) is found to be 100 %.

As inkjet printing is extremely attractive for fabrication on flexible substrates, an array of 20 $Gr_B/WS_2/Gr_T$ heterostructures were printed onto PET film on an area of 3 cm x 4 cm with no annealing performed, Figure 1d. The I-V curves in Figure 3a show non-ohmic behaviour under no or little illumination, with the similar power law observed for the device in Figure 2. Figure 3b shows that no photo-current saturation is reached for laser powers up to ~3 mW (higher powers cause substrate degradation), in contrast to the device in Figure 2. This is due to the thicker $Gr_T$ used in the case of PET substrate, which decreases the effective laser power reaching $WS_2$ . Bending test shows that photocurrent is stable up to ~2 % strain (Figure 3c), in agreement with previous results reported for heterostructures made by inks.[12] Note that at higher strain, the silver contacts break down. It is also interesting to compare the conductivity of the devices in Figure 2 and 3a: the out-of-plane resistance is 3.8 MΩ at 1V for the device in Figure 2, although the exact value depends on the voltage at which the current is measured due to the non-linearity of the I-V curve (see also Figure 4e). The in-plane resistances range from 250kΩ for $Gr_B$ to 600 kΩ for $Gr_T$: the larger resistance of $Gr_T$ is likely due to the more uneven surface on which the top contact is printed (i.e., on top of $Gr_B$ and $WS_2$). Both values include contact resistance (~ 10 kΩ). For the device in Figure 3a, the out-of-plane resistance is 13 MΩ at 0.5 V. Figure S14 shows the narrow distribution of out-of-plane conductivities for an array of heterostructures printed on glass.

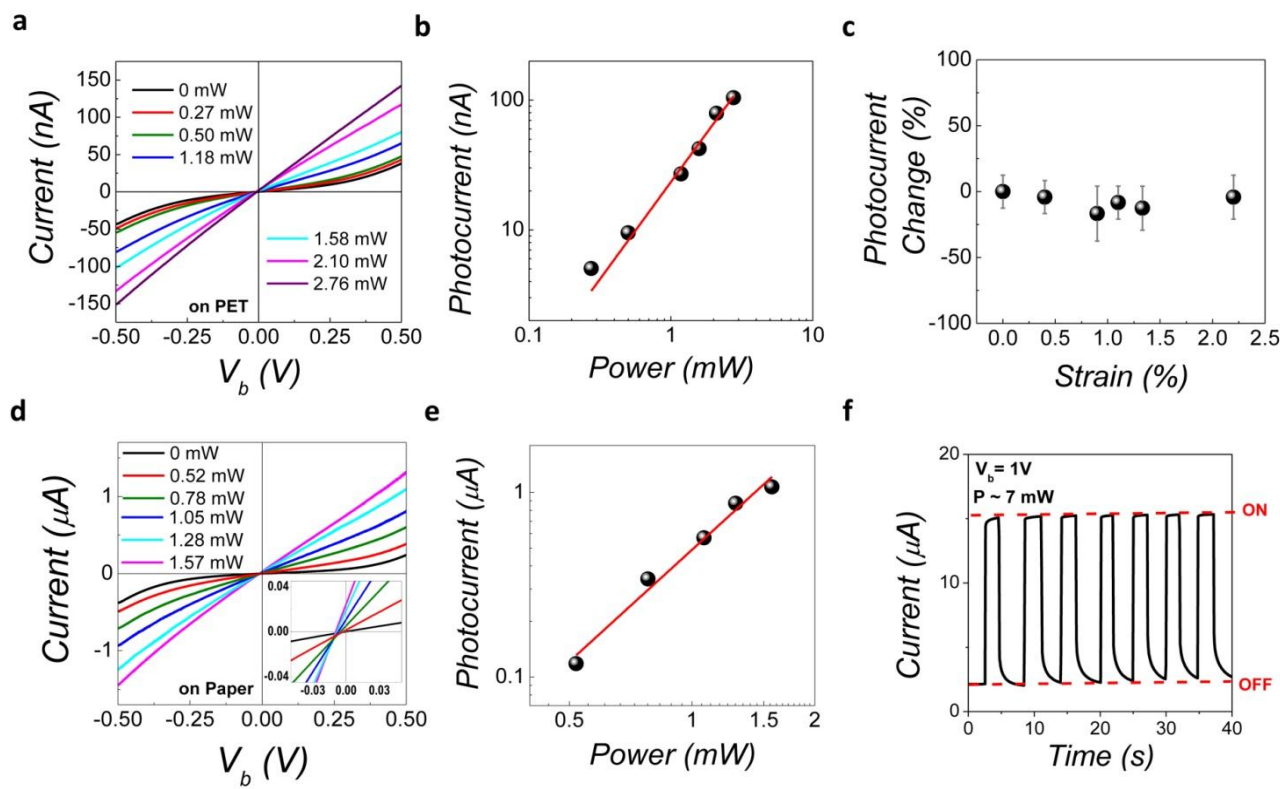

**Figure 3 | Flexible heterostructures fully printed onto plastic and paper. a**, I-$V_b$ curves of an all-printed $Gr_B/WS_2/Gr_T$ heterostructures on PET as a function of increasing laser power. (λ = 514 nm) **b**, photo-current measured at $V_b$ = 0.5 V and λ = 514 nm as a function of increasing laser power (the red line is an exponential fit of the experimental data). **c**, Bending test showing that the photo-current of an all-printed $Gr_B/WS_2/Gr_T$ heterostructures on PET is stable up to ~2% strain. **d**, I-$V_b$ curves of an all-printed $Gr_B/WS_2/Gr_T$ heterostructures on paper as a function of increasing laser power (λ = 514 nm). **e**, photo-current measured at $V_b$ = 0.5 V and λ = 514 nm as a function of increasing laser power (the red line is an exponential fit of the experimental data). Inset: zoom on the data for $V_b$ between -0.07 V and 0.07 V to show the changes in the short-circuit current with the laser power. **f**, Current generated by switching on and off the laser source and under $V_b$ = 1 V. Photoresponsivity higher than 1 mA/W is obtained.

We remark that several strategies can be used for improving the device performance. For example $Gr_B$ or $Gr_T$ can be replaced by Chemical Vapour Deposited (CVD) graphene, lowering the total resistance and increasing the optical transparency of the heterostructure. Figure S12 in the Supplementary Information shows $WS_2$ printed onto CVD graphene on silicon (Si/SiO$_2$): the inks form uniform patterns on the CVD graphene, no coffee ring or re-dispersion is observed, showing that inkjet printing and CVD technologies are perfectly compatible. A simple way to improve the device performance is to reduce the graphene electrode resistance. This can be easily obtained by maximizing the amount of material printed per unit area, either by increasing the ink concentration or by increasing the number of passes (see $Gr_B/MoS_2/Gr_T$ heterostructure, Section 4 in the Supplementary Information). In the case of porous substrates, such as paper, the solvent absorption allows minimizing the drop spacing during printing, which in turn increases the amount of material printed per unit area (Figure 1 f, blue points). Following this strategy, we fabricated a $Gr_B/WS_2/Gr_T$ photodetector on paper by printing a graphene ink with concentration of 2 mg/mL and using ~25 μm drop spacing, and 20 printing passes (Figure S15 in the SI). Figure 3 d shows the I-$V_b$ curves of the photosensor: the resistance decreases with increasing laser power. Figure 3 e shows the photo-current measured at Vb = 0.5 V and λ = 514 nm as a function of increasing laser power. Figure 3 f shows the current generated by switching on and off the laser every 10 seconds at $V_b$ of 1V. There is 1 order of magnitude increase in current, when switching from off to on, giving rise to responsivity higher than 1 mA/W, well above the typical resonsivity reported for devices made with liquid-phase exfoliated 2D crystals.[12,16] Note that for our devices a higher laser power can be used due to the increased thickness of the top graphene electrode.

On the other hand, the ability to design heterostructures and to make arrays by inkjet printing can be exploited to make novel devices. Here we show for the first time programmable logic memory devices, made completely with 2D crystals. Information storage is essential in any data processing system. As a consequence, while pursuing the goal of obtaining an "all inkjet" printed circuit, we have to necessarily address the issue of implementing such an important building block.

To this purpose, we propose a programmable logic memory device, enabled by the ink-jet printing technology, in which programming is performed at the time of fabrication. A "word" is stored in the memory through the definition of a horizontal stripe ("word-line") and vertical stripes, one for each bit of the word (which will be referred to as "bit-lines"), all of them made of graphene (Fig. 4a). A logic "1" is stored by short-circuiting the bit line to the word line, while a logic "0" is encoded by including a semiconducting layer (i.e., $WS_2$), between the word line and the corresponding bit line, which eventually suppresses the current. In Fig. 4b, we show the fabricated device, including the schematic of the complete circuit set up for the measurements, with the external bias voltage source $V_p$ and the load resistors $R_L$. In particular, when a voltage is applied to the word line (i.e., 0.5 V), the stored word is read in terms of the voltage $V_{Bi}$ across each pull-down resistor $R_L$, with higher voltage values (greater than 0.35 V) interpreted as logic "1" and lower voltage values (less than 0.135 V) as logic "0" (black solid line in Fig. 4c): in this specific case, the device is programmed with the word "010010001". We have also performed circuit simulations, considering a distributed resistance for the graphene word and bit lines, and assuming that the graphene/$WS_2$/graphene junction (logic "0") behaves as an ambipolar device. In particular, the I-V characteristics have been extracted from the measurement performed on the circuit in Fig. 4a, and included into the SPICE circuit simulator (in Fig. 4e we show an I-V characteristic of a junction as well as that of a short-circuit). Simulation results of the equivalent circuit of Fig. 4a (i.e., Fig. 4d) are shown in Fig. 4c (red dashed lines), showing good agreement between experimental and theoretical results.

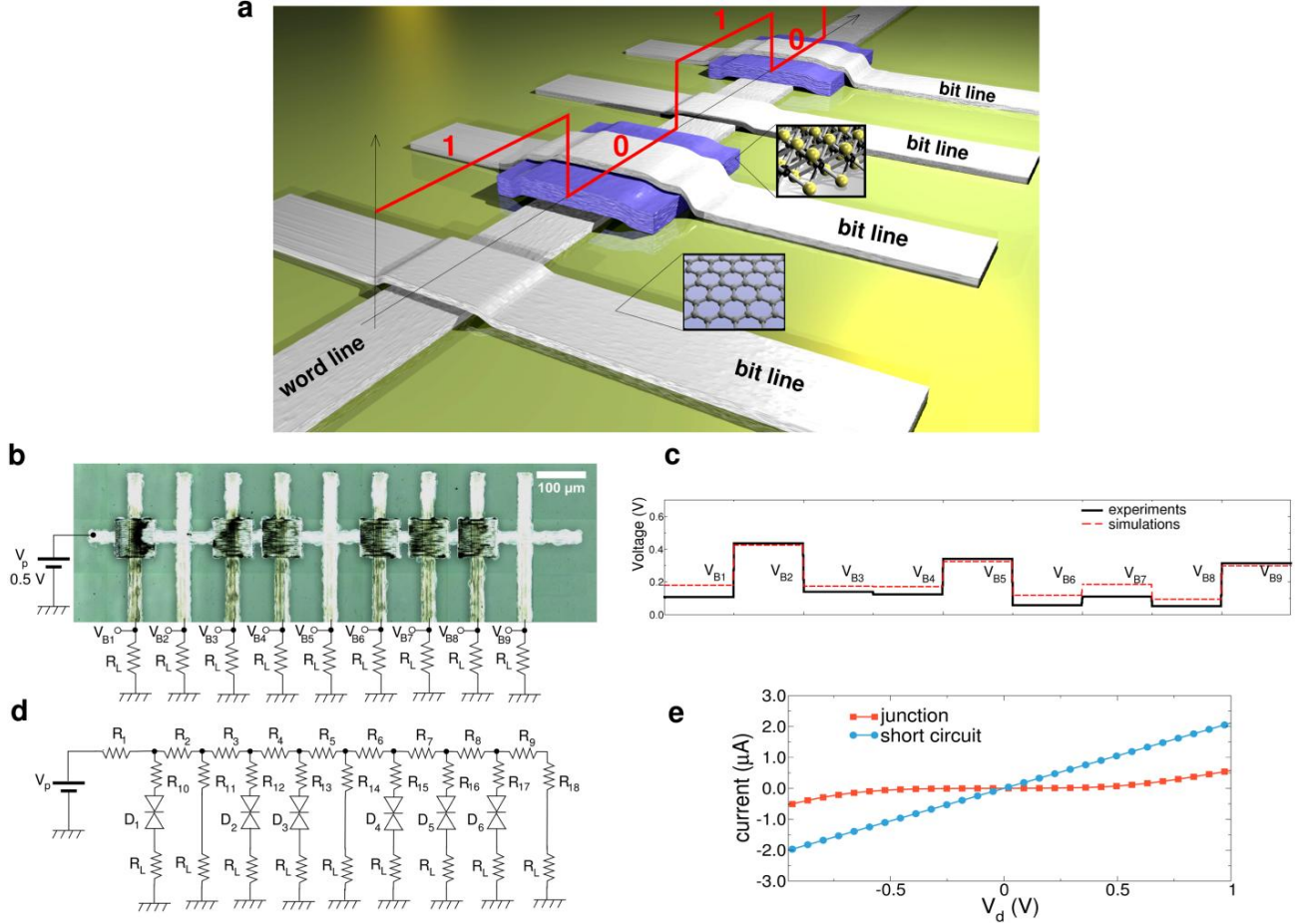

**Figure 4 | Logic memory device. a,** Sketch of the fabricated device – The Programmable Read Only Memory (PROM) is composed of a horizontal (word line) and vertical lines (bit lines) made of ink-jet graphene. A logic "1" is stored at regular intersections of the word line and the bit line, while a logic "0" is programmed by printing $WS_2$ between the two. The sketch shows a 4-bit memory storing the word "1010". **b,** Micrograph picture of the fabricated device, where the bias voltage ($V_p$) source and load resistors are added. **c,** Experimental (black solid line) and simulated results (red dashed line) for the operation of the circuit in a). **d,** Schematic of the equivalent electronic circuit. **e,** I-V characteristics of a graphene/$WS_2$/graphene junction and of a short-circuit.

The proposed device represents the first attempt to fabricate a memory with inkjet technology based on 2D materials and it can be exploited in a wide range of applications. As an example, it could be included in printed RF product tags, in order to store an identification number. Note that the inclusion of additional elements in the device, such as diodes and transistors, would allow the device to perform more complex functions and to increase the size of the read-only memory up to values of practical interest.

Finally, the potential adoption of 2D-crystal inks in displays, smart packaging and textiles, printed biological, chemical and environmental sensors as well as energy devices, requires early determination of the risk associated with exposure of living organisms. Furthermore, graphene is a remarkable material for biomedical applications.[39] To offer an initial indication of the biocompatibility profile of these water-based inks, we conducted a series of cytotoxicity studies. We studied the response to dose-escalated ink exposure using cell culture models representative of the human tissues that constitute the primary physiological barriers in non-occupational (e.g. skin exposure of consumers) and occupational (e.g. pulmonary exposure of workers production lines of such materials) scenarios. The viability of human lung (human alveolar epithelial cells; A549) and skin (human keratinocytes; HaCaT) cells was assessed using two assays: the modified LDH assay – a colorimetric assay used to evaluate cellular membrane damage induced by the material; and PI/Annexin V staining used to distinguish cell viability by flow cytometry (FACS).

The modified LDH assay was developed[40] to avoid potential interferences that are often reported as a result of interfering interactions between nanomaterials and reagents in colorimetric assays.[41] No differences in cell survival after 24hr of treatment with increased doses of the 2D inks were observed compared to untreated cells (Figure 5a). The LDH assay data was validated using FACS (Figure 5b and c) by staining with cellular markers of apoptosis (Annexin V) and necrosis (Propidium Iodide). Cells exhibiting different responses to 2D ink exposure were gated according to: healthy (unlabeled) cells (P2), early apoptotic (P3), late apoptotic and/or necrotic (P4) and necrotic cells (P5). Even at the highest dose of 100 µg/mL, cells appeared unstained, indicating the presence of predominantly alive cells (more than 90% of counted cells appeared in this region; Supplementary Information, Figure S16). Overall, no significant cytotoxic responses compared to untreated cells were observed for all doses (Supplementary Information, Figure S17). Optical microscopy of the 2D ink-treated cell cultures (50 µg/mL for 24 h, Figure 5 b and c) indicated strong interactions between the material and the cells that can lead to binding and internalization, to be further investigated in future studies. In the present work, no significant morphological changes indicative of cell death, such as decreased cellular confluence compared to the untreated cells; loss of cell/cell contact between neighbouring cells; contracted nuclei or multinucleated giant cells, were observed after treatment. Overall, the modified LDH, FACS and microscopy data all corroborated to indicate that the 2D material used did not induce any significant cytotoxic responses in the two cell lines and the dose-escalation regime studied here.

We remark that, to date, there have been very limited studies investigating the cytotoxic responses from exposure to exfoliated 2D materials that can inform about any biocompatibility limitations. The *in vitro* cytotoxic responses to $MoS_2$ nanosheets, using different exfoliating agents to the ones used in this study[42-44] have reported no significant adverse response using different cell lines (PC12, rat pheochromocytoma cells; Beas2B, human epithelial lung cells; THP-1, human myeloid cell line). In another study the cytotoxic responses of A549 cells using different lithium-based exfoliating agents have been reported.[45] Similarly, cytotoxic responses in HeLa (human cervical cancer cell line), 4T1 (mouse breast cancer cell line) and 293T (human embryonic kidney) cells to $WS_2$ nanosheets indicated that the type of exfoliating agent and their surface modifications play a critical role in the induced cytotoxicity.[46,47] In order to determine the biocompatibility profile of exfoliated 2D nanosheets thoroughly, more systematic studies should be performed, accounting for the anticipated route and dose of exposure and employing more complex models to determine potential adverse reactions. Here, we provided an initial comparative study between four types of 2D inks and two human cell lines that indicated biocompatibility within the dose escalation studied.

In conclusion, we have produced water-based, inkjet printable, and biocompatible 2D crystal inks that can provide new paradigms for manufacturing of fully printed 2D-crystal based devices of arbitrary complexity, which can be exploited in a wide range of applications. Due to the simplicity, and low cost of device fabrication and integration, we envisage this technology to find potential in smart packaging applications and labels, in particular for food, drinks, pharmaceuticals and consumer goods, where thinner, lighter and cheaper and easy to integrate components are needed.

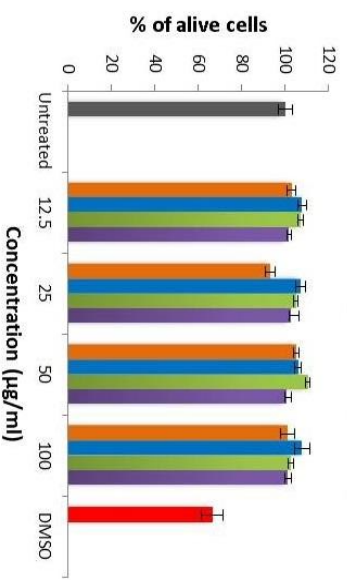
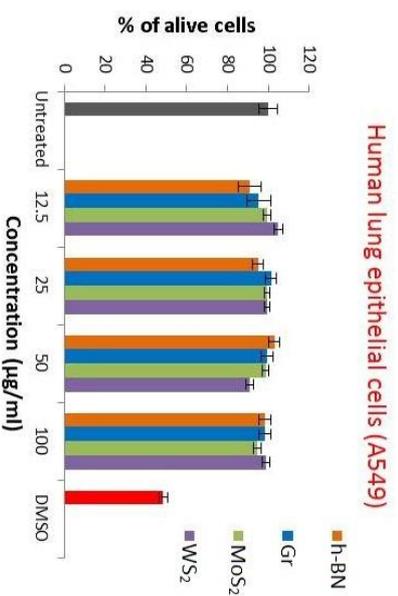
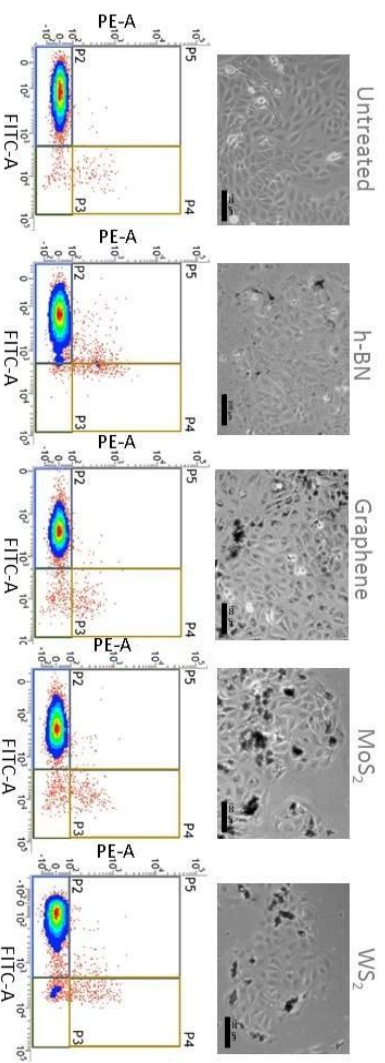
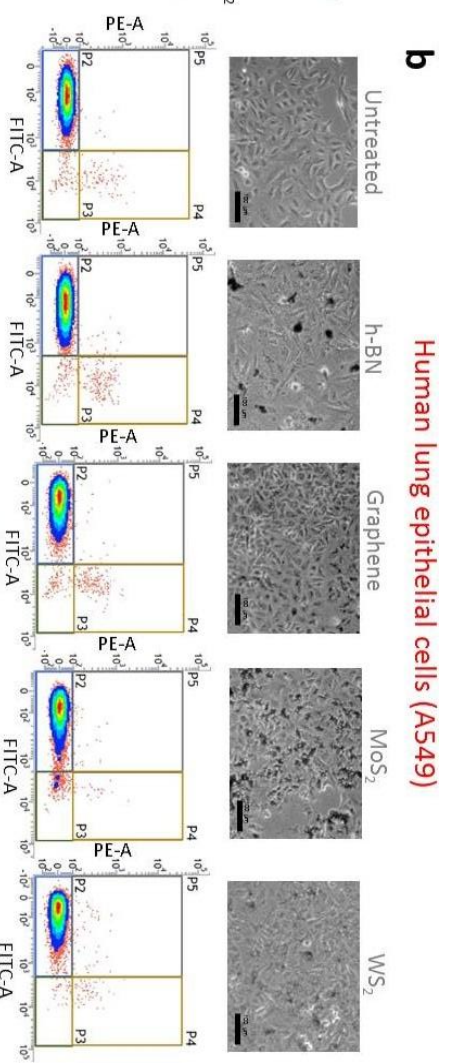

**Figure 5 | Cytotoxic responses of human lung (A549) and keratinocyte (HaCaT) cells to 2D crystals ink exposure. a,** Assessment of the toxicity of the material using modified LDH assay in A549 and HaCaT cells. Cells were treated at escalating doses of the material for 24h, lysed and centrifuged before the supernatant was carefully collected and incubated with the substrate for 15 min. Data are represented as mean ± SD (n = 6). **b,** and **c,** Top row: optical microscopy images of untreated and cells treated with 50 μg/ml boron nitride, graphene, molybdenum sulphide, tungsten sulphide inks for 24h. Scale bars on the images are 100 μm. Bottom row: assessment of the toxicity of the material by flow cytometry using PI/Annexin V staining. After incubation with the material for 24h, cells were harvested and labelled with Annexin V fluorochrome for 20 minutes. PI was added shortly before the analysis by flow cytometry to determine the % of unstained cells (alive cells, represented in the P2 region on the plots), Annexin V+/PI- cells (early apoptotic, represented in the P3 region), Annexin V-/PI+ cells (necrotic cells, represented in the P5 region) and Annexin V+/PI+ cells (late

**Methods**

**Ink Preparation.** Graphite (99.5% Grade) was purchased from Graphexel Ltd. All co-solvents, Triton x-100 and Xanthan gum were purchased from Sigma-Aldrich, UK. 10 µL Dimatix cartridges and PEL P60 paper were purchased from Printed Electronics Ltd. Si/SiO$_2$ wafers were obtained from IDB Technologies Ltd. CVD graphene was purchased from 2D-Tech Ltd. The dispersion was sonicated at 300W using a Hilsonic bath sonicator for 72 h. Flakes of the appropriate lateral size were collected via differential centrifugation to reduce the chance of nozzle blockages. The liquid obtained was centrifuged using a Sigma 1-14k refrigerated centrifuge at 903 g for 20 minutes before collecting the supernatant. The collected supernatant was then centrifuged at 16600 g for 1 hour and the precipitate re-dispersed in the printing solvent. The solvent consisted of less than 1:10 propylene glycol:water by mass, ≥0.06 mg / mL triton x-100 and ≥0.1 mg/mL XG. The concentration of 2D material was determined by UV-Vis spectroscopy, and the rheological properties were studied before printing (Supplementary Information, Sections 1 and 2). The lateral size of the nanosheets have been characterized by Atomic force Microscopy (Supplementary Information, Figure S5). No aggregation is observed when transferred in the cartridge or during printing. More details are provided in the Supplementary information.

**Printing.** We used a Dimatix DMP-2800 inkjet printer (Fujifilm Dimatix, Inc., Santa Clara, USA), which can create and define patterns over an area of about 200 x 300 mm and handle substrates up to 25 mm thick. A waveform editor and a drop-watch camera system allows manipulation of the electronic pulses to the piezo jetting device for optimization of the drop characteristics as it is ejected from the nozzle. The nozzle plate consists of a single raw of 16 nozzles of 23 µm diameter spaced 254 µm with typical drop size of 10 pL. Inks are printed onto a variety of substrates, in particular Silicon/SiO$_2$ (290 nm thermally grown oxide layer), SiO$_2$ (Quartz), PEL 60 paper (from Printed Electronics) and CVD graphene (from 2DTech). The printed features have been characterized by optical microscopy, Raman Spectroscopy, and Atomic Force Microscopy (Supplementary information, Section 3). Sheet Resistance has been measured with 2 contact probes, so the values include contact resistance.

**Photocurrent measurements**. Photocurrent maps were collected using a WiTEC Alpha 300 confocal Raman Microscope with 488 nm, 514.5 nm and 633 nm excitation wavelengths. Laser power was measured by a Thorlabs PM100D optical power meter. A bias voltage was applied using a Keithley 2614B Sourcemeter, also used to record I-V curves of the photoactive elements. The optically induced voltage change was measured across a 1 kΩ resistor by a Keithley 2182A Voltmeter and relayed into the WiTEC control software to generate a photocurrent map. National Instruments LabVIEW 2015 was used to control the sourcemeter and record data.

**Cell Culture.** Epithelial lung carcinoma cells (A549, ATCC, CCL-185) were maintained and passaged in Nutrient Mix F12 Ham media (Sigma-Aldrich, UK) supplemented with 10 % FBS (Thermo Scientific, UK), 50 µg/mL Penicillin, 50 µg/mL Streptomycin (Sigma-Aldrich, UK) at 37 °C in 5% $CO_2$. Cells were passaged twice a week using Trypsin-EDTA 0.05 % (Sigma-Aldrich, UK) when reached 80 % confluence. Activity of trypsin was stopped using 10 % FBS. Human keratinocyte cells (HaCaT) were generously provided by Dr Suzanne Pilkington (Dermatological Sciences, Institute of Inflammation and Repair, University of Manchester) and were maintained in DMEM media (Sigma-Aldrich, UK) supplemented with 10 % FBS, 50 µg/mL Penicillin, 50 µg/mL Streptomycin at 37 °C in 5 % $CO_2$. Cells were passaged twice a week using Trypsin-EDTA 0.05 % at 80 % confluence. Activity of trypsin was stopped using 10 % FBS.

**Cell Culture treatment.** Depending on the experiment, cells were seeded in 96 (LDH assay) or 12 (FACS analysis) well plates (Costar, Sigma) and treated when reached 70-80 % confluence. All treatments were performed in the cell culture medium in the absence of FBS, 10 % FBS was added to each well 4 h after the treatment. h-BN (1.6 mg/mL), Graphene (5.84 mg/mL), $MoS_2$ (1.44 mg/mL) and $WS_2$ (0.5 mg/mL) were vortexed shortly before making final dilutions for the treatment in the corresponding cell culture media. Cells were exposed to 12.5 - 25 - 50 - 100 µg/mL h-BN, Graphene, $MoS_2$ or $WS_2$ material for 24 h. Solvent (containing 0.06 mg/mL Triton x-100) used to stabilize the inks was also tested for its toxicity using the following concentrations: 0.012 – 0.006 – 0.003 – 0.0015 mg/ml Triton x-100 corresponding to the 100 – 50 – 25 – 12.5 µg/mL $WS_2$ inks.

**Annexin V-Alexa Fluor®488 conjugate/PI Assay.** After 24 h of treatment at indicated concentrations, supernatants were collected and cells were gently washed 3 times with PBS $Ca^{2+}/Mg^{2+}$ (Sigma-Aldrich, UK). Annexin-V staining was performed according to the instructions of the manufacturer (Thermo Fisher Scientific, UK). In brief, cells were trypsinized and merged with corresponding supernatants, centrifuged at 1500 rpm for 5 min, then re-suspended in 100 µL Annexin binding buffer (Thermo Fisher Scientific, UK) and stained with 2 µL Annexin V-Alexa Fluor®488 conjugate for 20 min at 15–25 °C. Propidium Iodide (1 mg/mL, Sigma) was added shortly before the analysis to the final concentration of 1.5 µg/mL. 10 000 cells were analysed on a BD FACSVerseTM flow cytometer using 488 nm excitation and 515 nm and 615 nm band pass filters for Annexin V and PI detection, respectively. Electronic compensation of the instrument was performed to exclude overlapping of the two emission spectra. Material alone was run in order to set up the gates including the cell population for the analysis. Percentage of unstained, cells stained with Annexin V, PI or both was calculated.

**Modified LDH Assay**. LDH assay was modified to avoid any interference coming from the interactions of the material with assay.[40] Briefly, the LDH content was assessed in intact cells that survived the treatment, instead of detecting the amount of LDH released in the media upon treatment. Media was aspirated and cells were lysed with 100 µL of lysis buffer for 45 min at 37 °C to obtain cell lysate, which was then centrifuged at 4,000 rpm for 20 min in order to pellet down the material. 50 µL of the supernatant of the cell lysate was mixed with 50 µL of LDH substrate mix (Promega, UK) in a new 96-well plate and incubated for 15 min at room temperature, after which 50 µl stop solution was added.

$$Cell\ Survival\ \% = (\alpha_{490nm}\ of\ treated\ cells/\alpha_{490nm}\ of\ untreated\ cells) \times 100 \quad (2)$$

The absorbance was read at 490 nm using a plate reader. The amount of LDH detected represented the number of cells that survived the treatment. The percentage cell survival was calculated using the equation above.

**Acknowledgments**

This work was partially supported by the Grand Challenge EPSRC grant EP/N010345/1, and the European Science Foundation (ESF) under the EUROCORES Programme EuroGRAPHENE (GOSPEL). F.W. acknowledges support from the Royal Academy of Engineering. DMM acknowledges the EPSRC in the framework of the NoWNano CDT. SS acknowledges support by the Army Research Office. SV would like to acknowledge the "RADDEL" project (Marie Curie Initial Training Network (ITN) grant number 290023 under the EU´s FP7 PEOPLE program. KK, GI and GF would like to gratefully acknowledge financial support from EU FP7-ICT-2013-FET-F GRAPHENE Flagship project (no. 604391). GI and GF gratefully acknowledge European Commission under Contract No. 696656 (Project 'GRAPHENE FLAGSHIP' Core 1). C.C. and D.M.M. acknowledge Kostya Novoselov, Steve Yeates, Joe Wheeler, and Adam Valentine Parry for useful discussions.

**Authors Contribution**

C.C. conceived and directed the experiments. D.M.M. developed the inks with inputs from V.S.R., and conducted all experiments, after preliminary results that were obtained by H.Y. and R.S., under the supervision of V.S.R and C.C. K.P. provided some electrical measurements on paper. S.S. did transfer the CVD graphene. Electrical characterization of the devices was performed by F.W. and D.M.M; G. F. conceived the logic memory device with inputs from C.C.; the device was fabricated by D.M.M and measured by M.M. The toxicology study was performed by S.V. under the supervision of K.K. The paper was written by C.C., D.M.M., G. F., S.V. and K.K. in close consultation with all authors.

**Additional information**

Competing financial interests: The authors declare no competing financial interests.

Supplementary Information contains additional details on the ink preparation and characterization of the inks, nanosheets, printed films and devices. Additional details on the biocompatibility tests and comparison with existing printable inks are also reported.